# Design of a Strong-Arm Dynamic-Latch based comparator with high speed, low power and low offset for SAR-ADC

Sounak Dutta
Electronics and Telecommunication
Jadavpur University
Kolkata, India
sounak04@gmail.com

***Abstract*—Comparators are utilised by Nyquist-rate and oversampling analog to digital converters (ADCs) to accomplish quantization and perhaps sampling. Thus, comparators have a substantial effect on the speed and accuracy of ADCs. This study provides a revised design for a dynamic-latch-based comparator that achieves the lowest latency, maximum area-efficient realisation, reduced power dissipation, and low offset. The proposed circuit has been designed and simulated using GDPK 45 nm standard CMOS-Process to operate on 100 MHz clock, at 1.2V supply voltage. Design and simulation have been carried out using CADENCE Virtuoso EDA tool. Compared to the original design, the PDP was easily reduced by approximately by 6% with offset voltage reduced by 8 mV without speed trade-off.**

## I. INTRODUCTION

Dynamic latched comparators are highly desirable for several applications, including high-speed analog-to-digital converters (SAR-ADCs) [1], memory sense amplifiers (SAs), and data receivers, owing to their high speed, low power consumption, high input impedance, and full-swing output. These days, dynamic comparators are widely being used because of their low-power consumption. Important design parameters for an ADC are fast speed, low offset, low power dissipation, and reduced chip area [2]. A comparator is a fundamental circuit that converts a signal from the analogue domain to the digital domain. It compares two voltage inputs and generates a binary signal indicating which one is greater. The output goes high if the non-inverting (+) input is higher than the inverting (-) input. The output goes low if the inverting input is higher than the non-inverting input. These days, dynamic comparators are widely being used because of their low-power consumption. In conventional dynamic latched comparators, a positive feedback mechanism (regenerative latch) is used which increases the speed of operation by instantly increasing a small-scale voltage difference at the input terminals to a full-scale digital level. In fact, in these comparators, usually there is no constant (current) path from the supply voltage to $\boldsymbol{V_{ss}}$, and they are controlled by a clock signal [3],[4]. However, the accuracy of these comparators is restricted by an input-referred latch offset voltage arising from threshold voltage $\boldsymbol{V_{th}}$, current factor $\boldsymbol{\beta = \mu C_{ox}.(W/L)}$, and parasitic node capacitance and output load capacitance mismatches [5, 6,7]. Utilizing the pre-amplifier before the regenerative output-latch stage will result in a reduced input-referred latch offset voltage and reduced kickback noise [8]. However, owing to ongoing technological scaling, the preamplifier-based comparators experience lower intrinsic gain with decreasing drain-to-source resistance $\boldsymbol{r_{ds}}$ and high - power consumption for a large bandwidth [9].

Therefore, for low power application we prefer Strong-Arm Dynamic Latch based comparator [10]. In [10] the proposed Strong-Arm Dynamic Latch based comparator is modified from Charge Sharing Dynamic Latch Comparator [11] and another Strong-Arm Dynamic Latch based comparator [12] by providing lower power consumption than both [11] and [12], lower offset voltage and less delay than [11] and reduced real estate over the chip than [12]. Although [10] has significantly lower offset voltage than [11], still the offset voltage value is marginally on the higher side if we require accurate comparison without pre-stage amplifier.

In this paper, we proposed a modified Strong-Arm Dynamic Latch comparator for lower offset, lower delay and lower PDP value (Power-Delay product) and compared its performance with the topology of [10]. Section II discusses the technique and design parameters, Section III gives simulation results and compares the design to the referred design, and Section IV concludes the article.

## II. METHODOLOGY & DESIGN PARAMETERS

***A. Basic Operation***: A comparator detects a differential input and produces a logical output based on the input difference's polarity.

The "Strong-Arm" comparator is made up of a regenerative latch pair added on top of a input clocked differential pair. As the cross-coupled latch pairs make strong positive feedback [10], it is able to make decisions quickly. The main advantage of Strong-Arm comparator is 1) it consumes zero static power, 2) it directly produces rail-to rail outputs, and 3) its input-referred offset arises from primarily one differential pair, so the offset voltage value is also lower.

***B. Design***: The proposed design of Strong-Arm Dynamic Latch based Comparator is shown in Fig.1. The schematic of [10] has been upgraded to operate with lower offset voltage by connecting two extra pMOS, $S_1$ and $S_2$ which is acting as pre-charge switches [13]. These two pre-charge switches also help to reset the output faster than the reset time measured in [10]. This schematic also makes use of a single, suitably sized clocked nMOS as tail transistor throughout the evaluation stage, which slightly decreases the evaluation delay than that of [10].

Switches $S_1$ and $S_2$ serve two purposes: a) they eliminate the prior states at nodes P and Q, therefore reducing dynamic offsets; and b) they provide a starting voltage of $V_{DD}$ at these nodes, so enabling amplification before $M_1$ and $M_2$ approach the triode area.

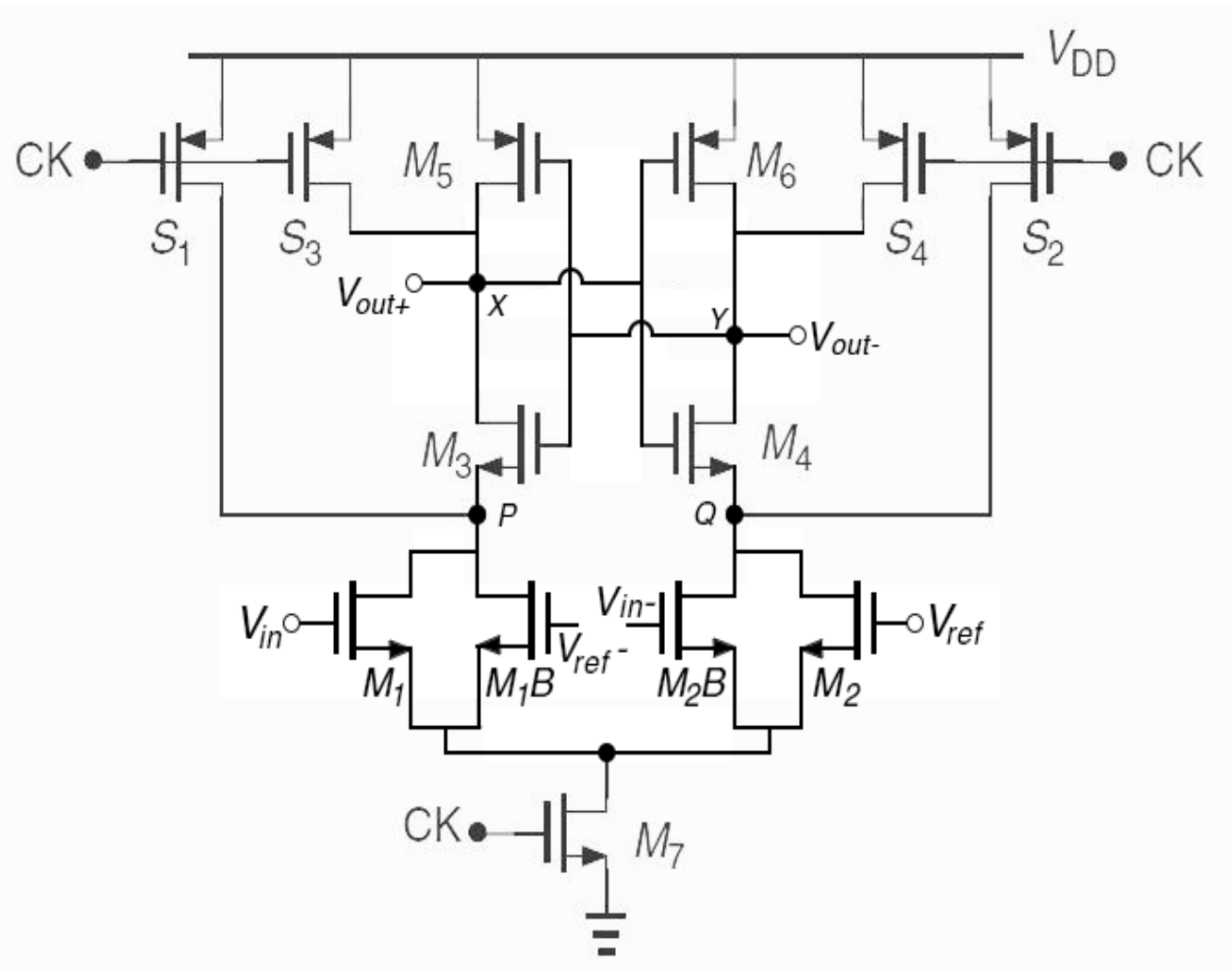

Fig 1.

***Operation of proposed comparator***

**Phase 1:** When CLK=0, the comparator enters the "Reset" phase, during which $S_1$- $S_4$ switches are activated and P, Q, X, and Y nodes in fig.1. are reset to $V_{DD}$. The parasitic capacitances of $\boldsymbol{C_P}$, $\boldsymbol{C_Q}$, $\boldsymbol{C_X}$, and $\boldsymbol{C_Y}$ are therefore charged to $V_{DD}$ during this phase. In Reset mode, $M_1$-M6 and $M_7$ transistors are cut off..

**Phase 2:** When CLK= 1, the comparator enters the Evaluation phase. In this state, switches $S_1$- $S_4$ are deactivated and the tail transistor $\mathbf{M_7}$ is in the triode region. Due to the difference in input voltage between $\boldsymbol{V_{in1}}$, $\boldsymbol{V_{in2}}$, the input pair transistors of $M_1$, $M_2$ are in the saturation zone. The pre charged capacitors $\boldsymbol{C_P}$ , $\boldsymbol{C_Q}$ are discharging at somewhat different rates. This indicates that the differential input voltage of the comparator is amplified by $M_1$ and $M_2$ and is reflected in the differential drain currents. In other words, this phase might result in voltage gain. This phase is known as the amplification mode. As the tail current is rather consistent over this period, we may write $|\boldsymbol{V_P} - \boldsymbol{V_Q}| \approx (\boldsymbol{g_{m1.2}}|\boldsymbol{V_{in}} - \boldsymbol{V_{ref}}|/\boldsymbol{C_{P,Q}})\ \boldsymbol{t}$, where $g_{m1.2}$ is the small-signal transconductance of $M_1$ and $M_2$.

**Phase 3:** When, the voltage on $C_p$ and $C_Q$ fall to $V_{DD}$ -$V_{THN}$ , $M_3$ and $M_4$ transistors turn on and $C_X$ and $C_Y$ begin to discharge. When the voltage on $C_X$, $C_Y$ decreases by a minimum of $V_{THP}$, cross-coupled latch transistors switch on and regenerate the comparator result; at this point, the lower voltage between X and Y nodes rapidly decreases to 0 and the other node is connected to $V_{DD}$ as shown in (Fig.2) and also explained in [13]. This comparator is completely dynamic and never draws static current due to transistors $M_3$ and $M_4$ [13,14].

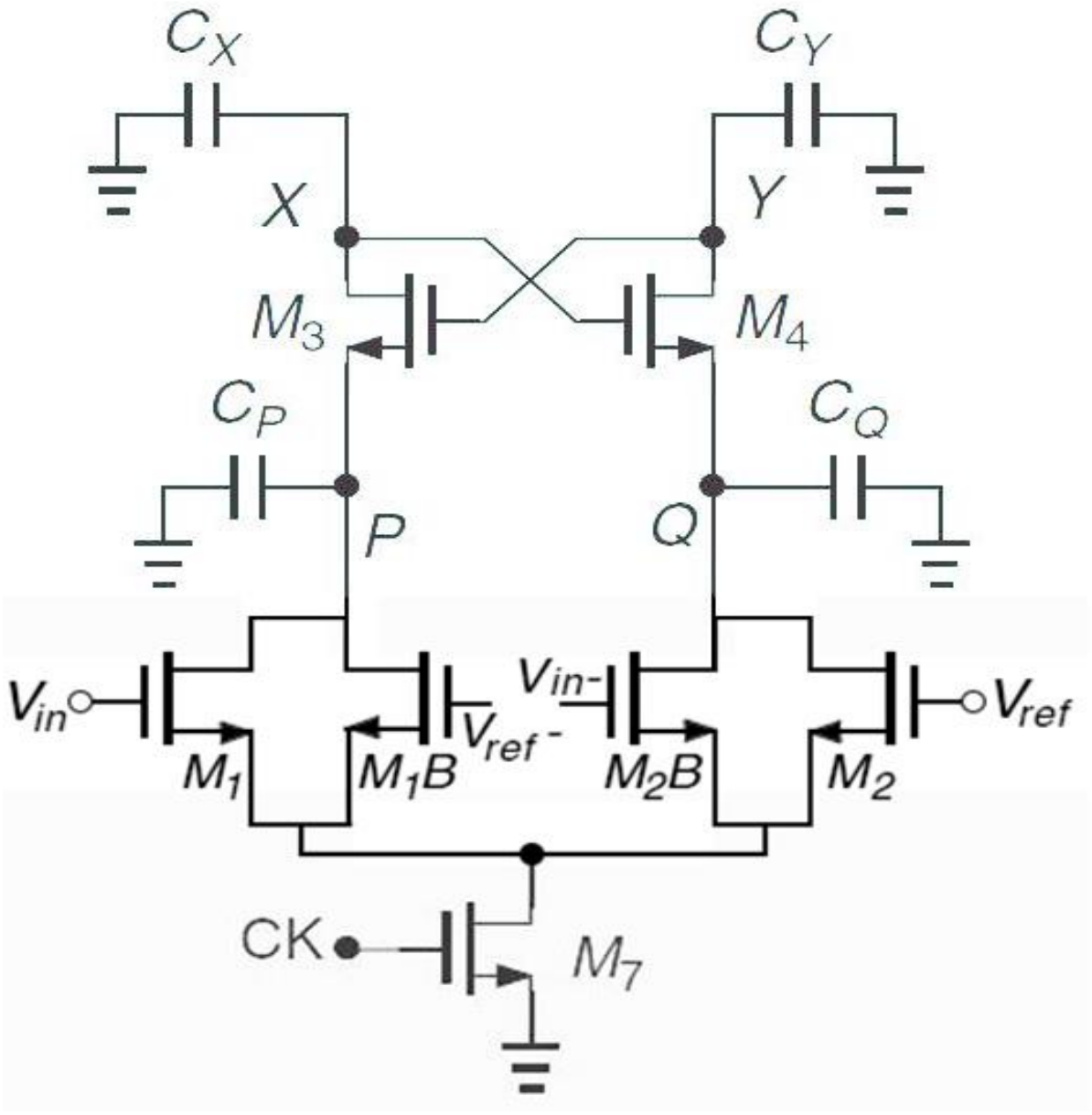

Fig.2.

***2) Delay Analysis***

The delay of the comparator consists of two components, $\boldsymbol{t_i}$ and $\boldsymbol{t_{reg}}$ (latch delay). The first term, $\boldsymbol{t_i}$, represents the time taken by the load capacitor to discharge until the first pMOS transistor becomes ON.

As described in [10], if $\boldsymbol{V_{in1}} > \boldsymbol{V_{in1}^-}$ transistors $M_1$ and $M_3$ transistors produce a quicker discharge of $\boldsymbol{V_{out}^-}$ and activate $M_6$. On this basis, the delay may be calculated as follows:

$$\boldsymbol{t_i} = \frac{C_{load}}{I_{D1}} \cong 2.\frac{C_{load}.V_{THP}}{I_{tail}} \qquad (1)$$

In (1), for minimal input differential voltage ($\Delta\boldsymbol{V_{in}}$), the drain current may be estimated as constant and it is equivalent to half the tail current. The second term $\boldsymbol{t_{reg}}$ , represents the entire latching delay of two cross coupled inverters. It is estimated that the final output will be half of the supply rail ($\Delta\boldsymbol{V_{out}} = \boldsymbol{V_{DD}/2}$) depending on $\Delta\boldsymbol{V_0}$ ,the primary voltage differential [10].

The equation for latch evaluation delay $\boldsymbol{t_{reg}}$ as given in [15,16] is dependent on the primary voltage difference during the commencement of the regeneration phase.

$$\boldsymbol{t_{reg}} = \frac{C_{load}}{g_{m(eff)}}.\ln\left(\frac{\Delta V_{out}}{\Delta V_0}\right) \cong \frac{C_{load}}{g_{m(eff)}}.\ln\left(\frac{V_{DD}/2}{\Delta V_0}\right) \qquad (2)$$

In the above equation $\boldsymbol{g_{m(eff)}}$ is the effective trans-conductance of cross-coupled inverters [12].

We know, $\Delta\boldsymbol{V_0}= |\boldsymbol{V_{out}^+} - \boldsymbol{V_{out}^-}|_{t=t_0}$. Now by using equation (1) we get,

$$\Delta\boldsymbol{V_0} = |\boldsymbol{V_{THP}}| - \frac{I_{D2}.t_i}{C_{load}} \qquad (3)$$

The difference in input current $\Delta I_{in}$ between the two branches is considerably less than the individual currents $I_{D1}$ and $I_{D2}$ , which can be equivalent to half of $I_{tail}$.

So, now (3) becomes,

$$`\Delta V_0 = |V_{THP}| \left(\frac{\Delta I_{in}}{I_{D1}}\right) \cong 2.\, |V_{THP}| \left(\frac{\Delta I_{in}}{I_{tail}}\right)$$

By solving this we get,

$$\Delta V_0 = 2.\, |V_{THP}| \left(\frac{\Delta I_{in}}{I_{tail}}\right) \sqrt{\frac{\beta_{1.2}}{I_{tail}}}\, .\Delta V_{in} \qquad (4)$$

Here, $\beta_{1.2} = \mu C_{ox} \, (W/L)_{1,2}$ is the gain factor in $\frac{\mu A}{V^2}$.

$\beta_{1.2}$ represents the input transistor's current factor. The tail current $I_{tail}$ is a function of the supply voltage and the input common mode voltage. By substituting $\Delta V_0$ (from (4)) into (2) and the value of $t_i$ from (1), the total delay may be calculated as illustrated in (5).

$$t_{total} = t_i + t_{reg}$$

$$t_{total} = 2.\frac{C_{load}.V_{THP}}{I_{tail}} + \frac{C_{load}}{g_{m(eff)}}.\ln\left(\frac{V_{DD}/2}{\Delta V_0}\right) \qquad (5)$$

Equation of (5) can be written as:

$$t_{total} = 2.\frac{C_{load}.V_{THP}}{I_{tail}} \frac{C_{load}}{g_{m(eff)}}\, .\ln\left(\frac{V_{DD}/2}{2.|V_{THP}|\left(\frac{\Delta I_{in}}{I_{tail}}\right)\sqrt{\frac{\beta_{1.2}}{I_{tail}}}\, .\Delta V_{in}}\right)$$

(6)

This is the total analytical delay of the proposed dynamic latch comparator. [10]

### 2) *Power calculation*

Basically, to determine the average power wasted from the supply voltage throughout a single comparison time, we use the below equation:

$$P_{avg} = \frac{1}{T}\int_0^T V_{DD}\, .I_D dt \qquad (7)$$

Where T is the period of the comparator clock signal and the total current drawn from supply voltage is denoted by $I_D$. So the above formula can be written as

$$P_{avg} = f_{CLK}\int_0^T V_{DD}\, .I_D dt = f_{CLK}.V_{DD}\int_0^T I_D dt \qquad (8)$$

### 3) *Offset Voltage*

The input offset voltage of a comparator is the input voltage at which its output changes from one logic level to the other. It may be caused by device mismatch or may be inherent to the design of a comparator. In other words Offset is the input error range below which the comparator is unable to identify the specified minimum voltage difference. As a result, the comparator's resolution and speed are constrained.

In this circuit the offset voltage is reduced from that of [10] by introducing $S_1$ and $S_2$. In Fig. 1, the pre-charge action of $S_1$- $S_4$ keeps transistors $M_3$- $M_6$ off at first, minimising their offset contribution. Now when referenced to the input mismatches between $M_3$ and $M_4$ in a typical design are divided by approximately a factor of $A_V \approx 4$, and those between $M_5$ and $M_6$ by roughly a factor of 10 (since these transistors switch on only towards the end). As a result, $M_1$ and $M_2$ become the primary contributors [13].

By making use of the fact that any capacitive load difference at the point P and Q, ($\Delta C = C_P - C_Q$) induces a shift in the trip point shown by the following first-order equation of (9), the comparator input offset can be programmed.

$$V_{io} = \frac{I_{D1}}{g_{m1}}\frac{\Delta C}{C_P} = \frac{V_{ov1}}{2}\frac{\Delta C}{C_P} \qquad (9)$$

Here $V_{io}$ is the input offset voltage, $I_{D1}$ is the average current, $g_{m1}$ is the transconductance of $M_1$, $\Delta C$ is capacitive load difference at balanced state, and $V_{ov1}$ is the overdrive of $M_1$ and $M_2$ in saturation during the initial drain node discharging phase.

Offset cancellation can be performed by introducing binary-sized array of MOS capacitors on both sides of the comparator, with the ability to digitally change the value of any single capacitance shown in Fig. 3. Now in order to further reduce the offset node P and Q can be equipped with several small unit capacitors but will reduce speed and increase the power dissipation [13,17].

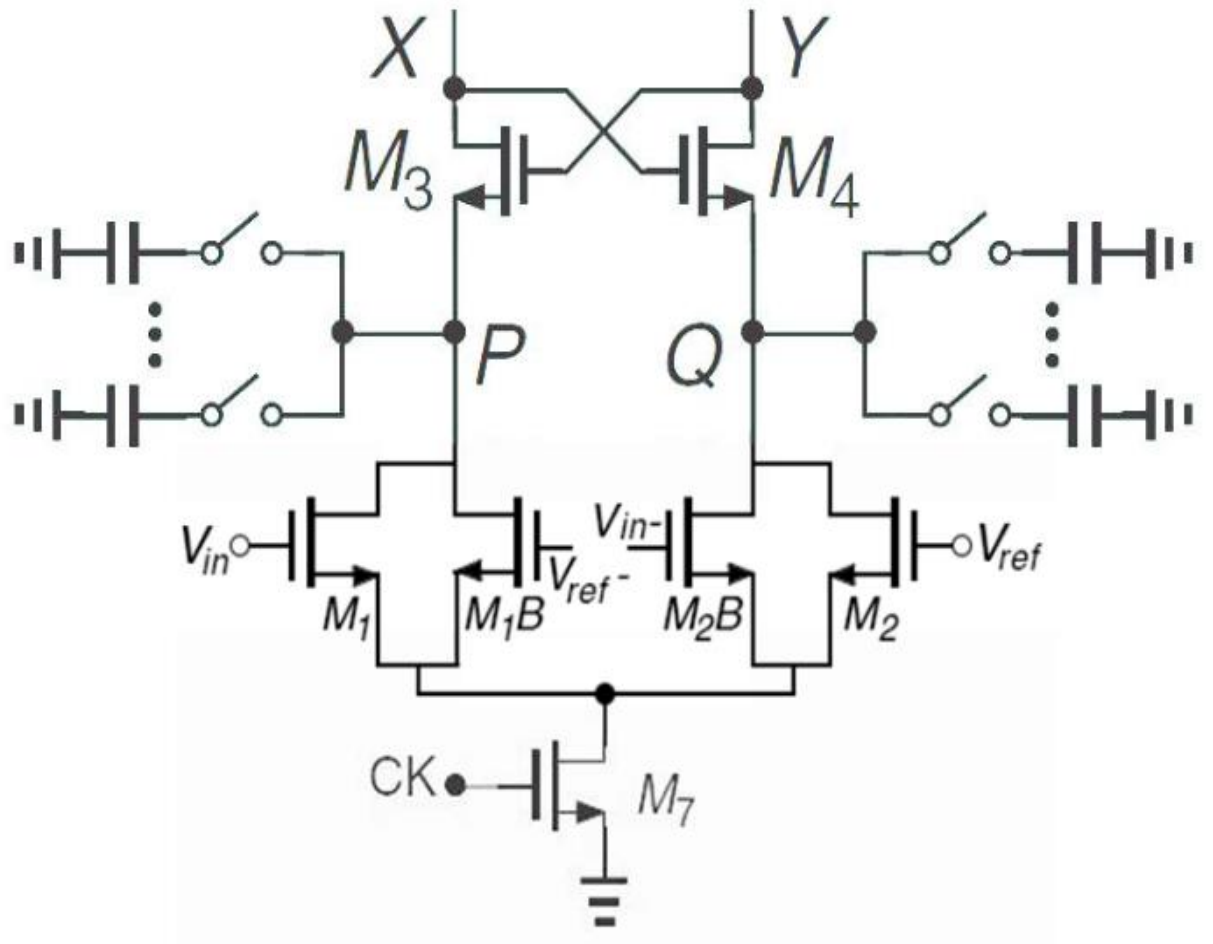

Fig. 3.

### *Transistor aspect ratio*

In order to begin the process of designing a comparator, the first thing that must be considered is to select a optimal transistor sizes that satisfies the offset requirement, provides less latency, and consumes less power. In this case, the size of the transistors have been optimized to be as low as feasible in order to reduce delay and minimise parasitic capacitance.

TABLE 1: TRANSISTOR ASPECT RATIOS (W, L ARE WIDTH AND LENGTHS OF TRANSISTORS)

| Transistor | [10] $W/L$ | Proposed circuit $W/L$ |
|---|---|---|
| $M_1$- $M_2$ | $180nm/45nm$ | $180nm/45nm$ |
| $M_{1B}$ – $M_{2B}$ | $180nm/45nm$ | $180nm/45nm$ |
| $M_3$- $M_4$ | $180nm/45nm$ | $210nm/45nm$ |
| $M_7$ | $180nm/45nm$ | $180nm/45nm$ |
| $M_5$- $M_6$ | $285nm/45nm$ | $120nm/45nm$ |
| $S_3$- $S_4$ | $285nm/45nm$ | $270nm/45nm$ |
| $S_1$- $S_2$ | - | $120nm/45nm$ |

## III. RESULTS AND DISCUSSIONS

The design is simulated on the Cadence Virtuoso platform using GDPK 45nm Technology with $V_{DD}$ = 1.2V. Throughout the procedure, the CLK frequency is set to 100MHz.

Topology of [10] is also simulated on the same platform. The exemplary screenshot of the simulated waveforms for the proposed system is depicted in Fig. 4(a) and 4(b). The same for the [10] has been removed to prevent and for the sake of concision. After comparing the results of the present design to that of [10], Table II lists the specifications and enhancements of the current design.

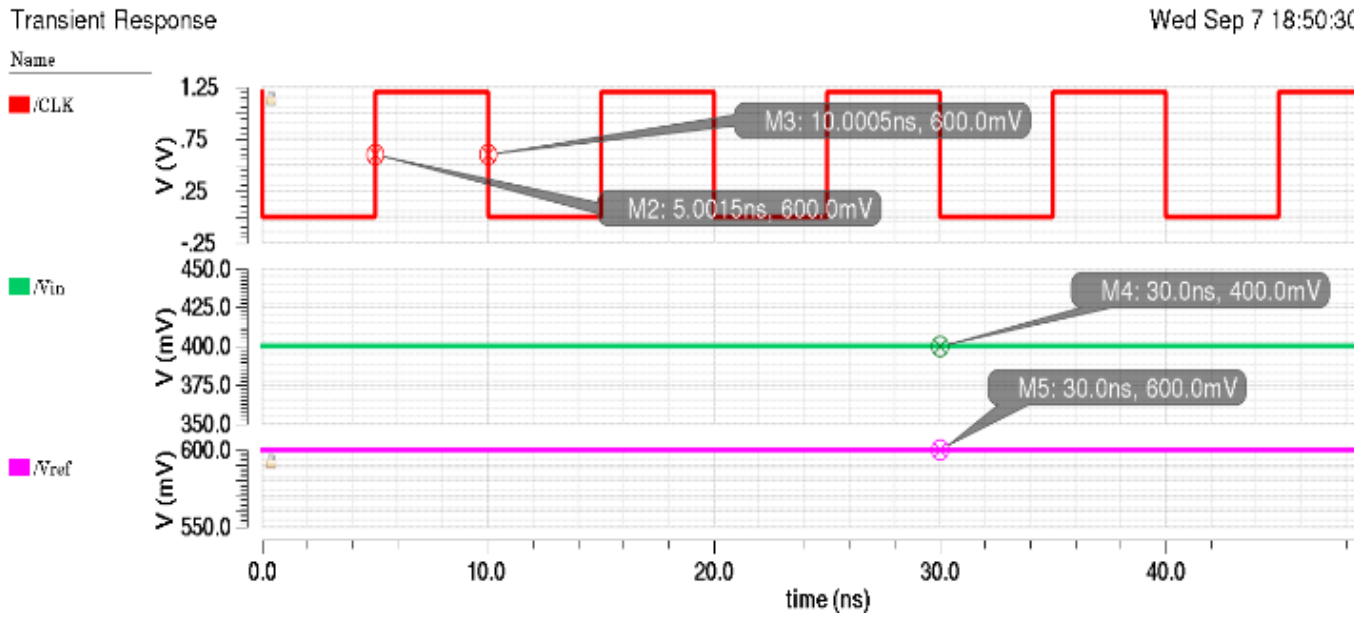

Fig. 4(a)

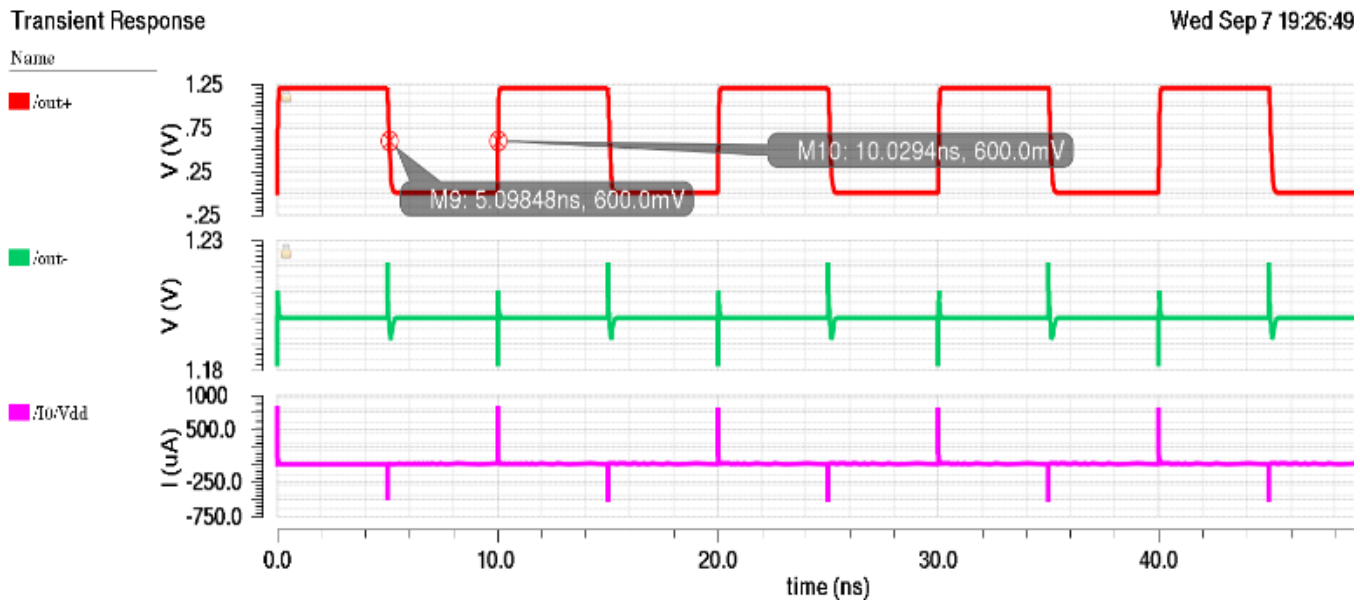

Fig. 4(b). Simulation results of proposed circuit for $V_{DD}$ = 1.2V, $V_{in}$= 400mV $V_{ref}$ = 600mV, CLK frequency = 100 MHz

Fig. 4(a) and 4(b)'s simulation results show that only one output, $V_{out}^{+}$ is changing states during the evaluation period (CLK= 1), while the other, $V_{out}^{-}$, remains high (at $V_{DD}$). This decreases average power consumption. According to Fig. 4(b), the pre-charge (reset) and evaluation periods are 28.9ps and 96.9ps, respectively. Therefore, the obtained average delay is 62.9ps. This value is less than that of [10]. which is one of the primary advantage of this circuit.

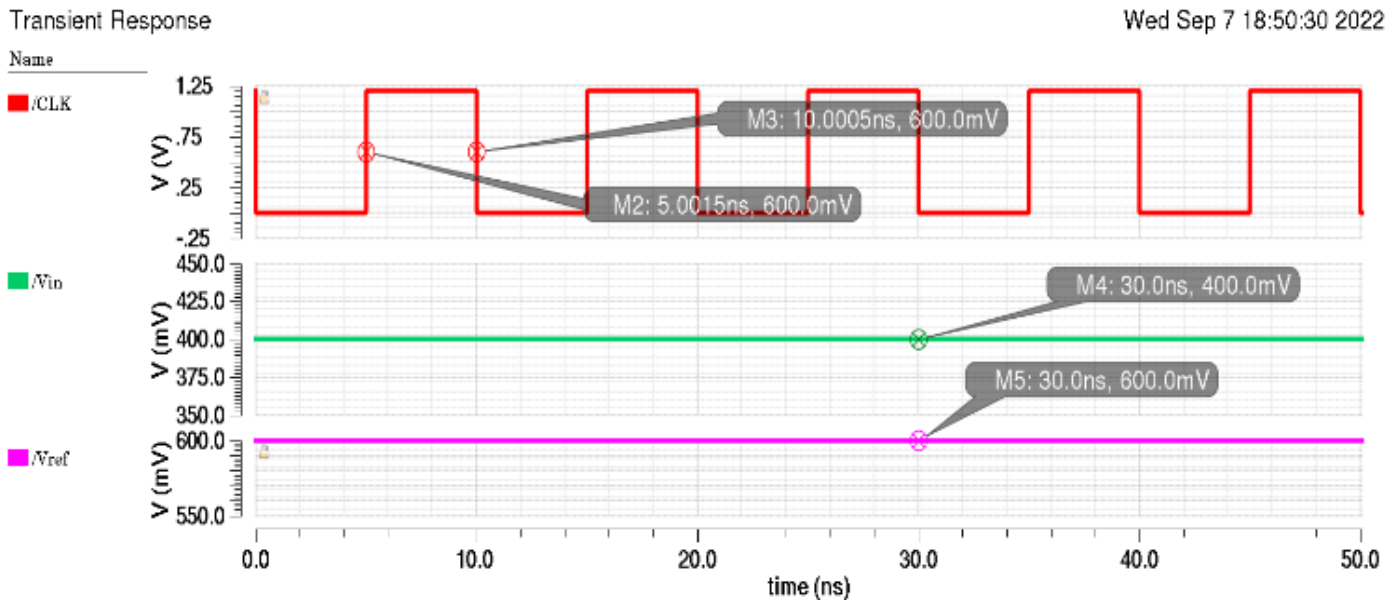

Fig 4(a)

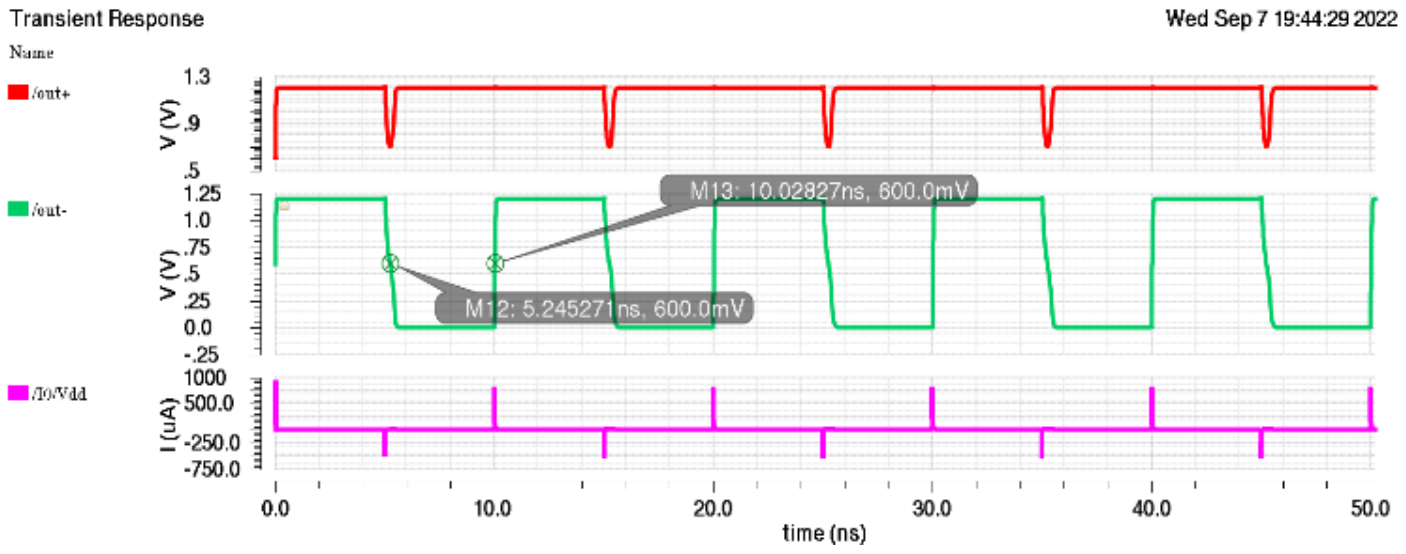

Fig. 4(c). Simulation results of proposed circuit for $V_{DD}$ = 1.2V, $V_{in}$= 400mV $V_{ref}$ = 300mV, CLK frequency = 100 MHz

Fig. 4(a) and 4(c)'s simulation results show that the pre-charge (reset) and evaluation times are set to be 27.8ps and 243.8ps, respectively. The evaluation time in the waveform in Fig. 4(c) is prolonged when compared to Fig. 4(b). Due to $V_{in}$ and $V_{ref}$'s proximity to one another, the comparator must compare more bits in order to make a judgement, which increases reaction time [10]. As a result, the average delay is 135.8ps. 4.2732 μW is the average power dissipation.

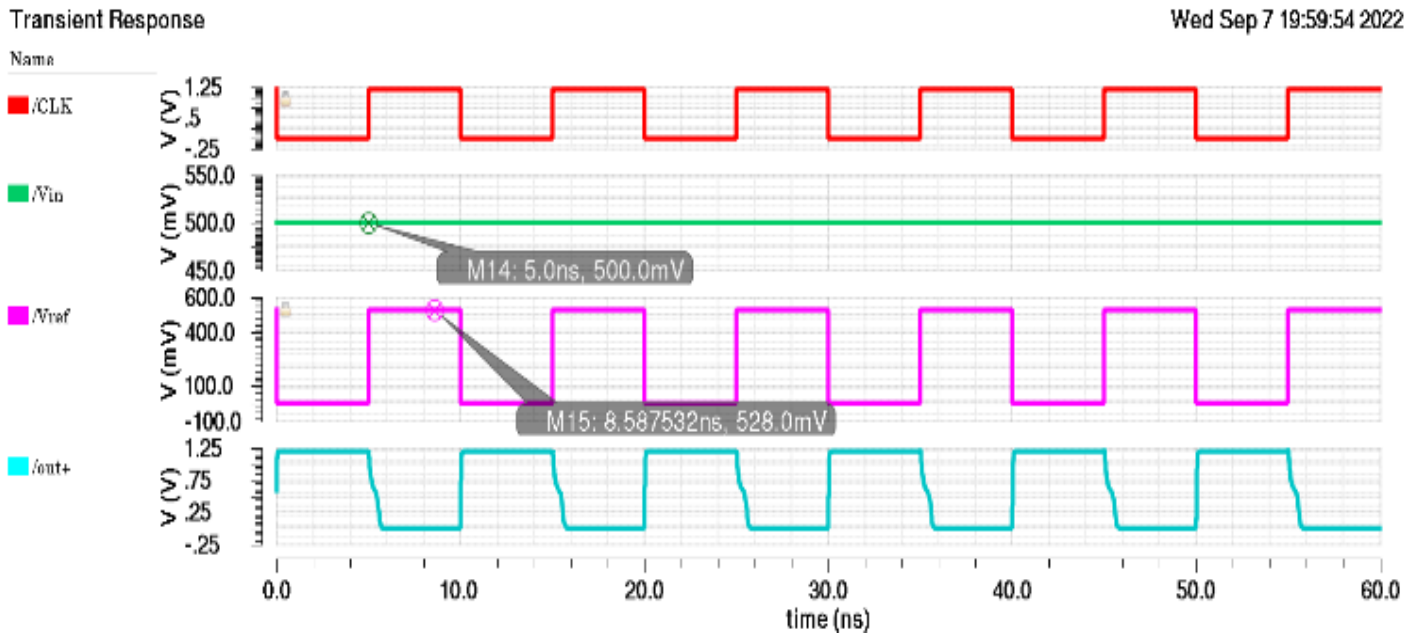

Fig. 5. Offset voltage of proposed design

The offset voltage measured from Fig. 5. is 28mV.

TABLE 2: PERFORMANCE COMPARISON

| Parameter / Topology | Average Dynamic Power($\mu$W) | Average Delay (ps) | PDP (fJ) | Offset Voltage(mV) |
|---|---|---|---|---|
| KasiBandla et al [10] | 4.237 | 67.43 | 0.286 | 36 |
| Proposed circuit | 4.272 | 62.9 | 0.268 | 28 |

## IV. CONCLUSION

In this paper design of a modified strong-arm dynamic latch comparator is presented that can operate at high frequency, consumes low power and has a low offset on a supply of $V_{DD}$ = 1.2V. This paper has been evaluated against [10]. Important design parameters are shown in TABLE 2. Both the original topology and proposed circuit topology have been simulated using CADENCE Virtuoso with GDPK 045nm technology. Table 1. Presents the optimized transistor sizes. In the proposed design delay is reduced by 6%. Although the power consumption slightly higher than that of [10] but overall PDP is also improved by nearly 6%. When comparing to [10], the proposed design offers a lower offset value without even a trade off with speed.

## ACKNOWLEDGMENT

Authors would like to thank staffs of IC Centre, Jadavpur University for their valuable support guidance.